\documentclass{PoS}

\leftline{\parbox{20cm}{\large DESY 12 - 181, Edinburgh 2012/20}}

\title{Perturbative subtraction of lattice artifacts in the computation of renormalization constants}

\ShortTitle{Perturbative subtraction of lattice artifacts in the computation of renormalization constants}

\author{M.~Constantinou$^a$, M.~Costa$^a$,  M.~G\"ockeler$^{b}$, R.~Horsley$^{c}$,  
H.~Panagopoulos$^a$,\newline
\speaker{H.~Perlt}$^d$,
   P.~E.~L.~Rakow$^e$, G.~Schierholz$^f$,
   A.~Schiller$^d$\\
\llap{$^a$}Department of Physics, University of Cyprus, Nicosia, CY-1678, Cyprus\\
\llap{$^b$}Institut f\"ur Theoretische Physik, Universit\"at Regensburg,
93040 Regensburg, Germany\\
\llap{$^c$}School of Physics, University of Edinburgh,
Edinburgh EH9 3JZ, UK\\
\llap{$^d$}Institut f\"ur Theoretische Physik, Universit\"at Leipzig,
04109 Leipzig, Germany\\
\llap{$^e$}Theoretical Physics Division, Department of Mathematical Sciences,
University of Liverpool, Liverpool L69 3BX, UK\\
\llap{$^f$}Deutsches Elektronen-Synchrotron DESY,
22603 Hamburg, Germany\\
\\
{\rm E-mail}: perlt@itp.uni-leipzig.de}

\abstract{The determination of renormalization factors is of crucial importance.
They relate the observables obtained on finite, discrete lattices to their 
measured counterparts in the continuum in a suitable renormalization scheme. 
Therefore, they have to be computed as precisely as possible. A widely used
approach is the nonperturbative Rome-Southampton method. It requires,
however, a careful treatment of lattice artifacts. 
They are always present because simulations are done at lattice spacings $a$
and momenta $p$ with $ap$ not necessarily small.
In this paper we try to suppress these artifacts by subtraction of one-loop
contributions in lattice perturbation theory. We compare results obtained from a complete one-loop subtraction
with those calculated for a subtraction of $O(a^2)$.}

\FullConference{The 30th International Symposium on Lattice Field Theory\\
		 June 24 - 29,  2012\\
		 Cairns, Australia}

\begin{document}

\section{Introduction}

Renormalization factors relate observables computed on finite lattices to their
continuum counterparts in specific renormalization schemes. Therefore,
their determination should be as precise as possible in order to allow for
a reliable comparison with experimental results. One approach is based
on lattice perturbation theory~\cite{Capitani:2002mp}. 
However, it suffers from its intrinsic
complexity, slow convergence and the impossibility to handle mixing
with lower-dimensional operators. Therefore, nonperturbative methods
have been developed in the last years. Among them the so-called
Rome-Southampton method~\cite{Martinelli:1994ty} (or RI-MOM scheme) is widely used because
of its simple implementation. It requires, however, gauge fixing.

In a recent paper~\cite{Gockeler:2010yr} some of us have given a comprehensive discussion
and comparison of perturbative and nonperturbative renormalization.  
One of the conclusions was the possibility
to suppress the unavoidable lattice artifacts by subtracting them
perturbatively. For simple operators this can be done in one-loop
order completely by computing the corresponding diagrams numerically.
While being very effective this procedure is rather involved and
not suited as a general method for more complex operators, especially
for operators with more than one covariant derivative.
A more general approach could be based on the subtraction of
one-loop terms of the order $a^2$ with $a$ 
being the lattice spacing. The computation of those
terms has been pioneered by the Cyprus group~\cite{Constantinou:2009tr}
and applied to various operators for different actions.

In this paper we apply this ``reduced'' subtraction procedure to
some exemplary operators and compare the results with those
of the complete one-loop subtraction as given in~\cite{Gockeler:2010yr}.
We investigate the point operators  $\mathcal{O}^{\rm S}=\bar{u}\,d,\,
\mathcal{O}_\mu^{\rm V} = \bar{u}\,\gamma_\mu\,d, \,\mathcal{O}_\mu^{\rm A}=\bar{u}\,\gamma_5\gamma_\mu\,d$
and $\mathcal{O}_{\mu\nu}^{\rm T}=\bar{u}\,\sigma_{\mu\nu}\,d$ for light quarks ($u, d$).
The corresponding Z factors have been measured 
(and chirally extrapolated) at $\beta=5.20, 5.25, 5.29$ and  $5.40$.
We used clover improved Wilson fermions with plaquette gauge action.
All results are computed in Landau gauge. The clover parameter $c_{SW}$
used in the perturbative calculation is set to its lowest order value $c_{SW}=1$.

\section{Renormalization group invariant operators}

In the RI-MOM scheme the renormalization constant $Z$ is obtained by imposing the condition
\begin{equation}
\frac{1}{12}\, {\rm tr} \left(\Gamma_R(p)\,\Gamma^{-1}_{\rm Born}(p)\right) = 1
\label{ZDet1}
\end{equation}
at $p^2=\mu^2$, where $\Gamma$ is the corresponding amputated Green function
of the operator $\mathcal{O}$ under study. The Z factors relate the renormalized to the unrenormalized
Green functions
\begin{equation}
\Gamma_R(p) = Z_q^{-1}\,Z \, \Gamma(p)\,,
\end{equation}
with $Z_q$ being the quark field renormalization constant determined by
\begin{equation}
Z_q(p)= \frac{{\rm tr}\left(-i \sum_\lambda \gamma_\lambda \sin (a p_\lambda) a S^{-1}(p)\right)}{12\sum_\lambda \sin^2(ap_\lambda)}\,,
\end{equation}
($S^{-1}$ is the inverse quark propagator).
Using (\ref{ZDet1}) we compute $Z$ from
\begin{equation}
Z_q^{-1}\,Z\, \frac{1}{12}\,{\rm tr}\left(\Gamma(p)\,\Gamma^{-1}_{\rm Born}(p)\right) = 1\,.
\label{ZDet2}
\end{equation}
For operators which carry
at least one space-time index (i.e. the corresponding $O(4)-$ multiplet
has dimension greater than 1) we use an averaging procedure as described
in~\cite{Gockeler:2010yr}.

We define the so-called renormalization group invariant (RGI) operator, which is independent
of scale $M$ and scheme $\mathcal{S}$, by~~\cite{Gockeler:2010yr}
\begin{equation}
\mathcal{O}^{\rm RGI} = \Delta\,Z^{\mathcal{S}}(M)\, \mathcal{O}^{\mathcal{S}}(M) = Z^{\rm RGI}(a) \,\mathcal{O}_{\rm bare}\,
\label{RGI1}
\end{equation}
with
\begin{equation}
 \Delta\,Z^{\mathcal{S}}(M) = \left(2\beta_0 \frac{g^\mathcal{S}(M)^2}{16\,\pi^2}\right)^{-(\gamma_0/2\beta_0)}
 \, {\rm exp}\left\{ \int_0^{g^{\mathcal{S}}(M)} dg'   \left( \frac{\gamma^{\mathcal{S}}(g')}{\beta^{\mathcal{S}}(g')}+
 \frac{\gamma_0}{\beta_0 g'} \right)  \right\}
 \label{RGI2}
\end{equation}
and
\begin{equation}
Z^{\rm RGI}(a) = \Delta\,Z^{\mathcal{S}}(M) \, Z^{\mathcal{S}}_{\rm bare}(M,a)\,.
\label{RGI3}
\end{equation}
$g^\mathcal{S}$, $\gamma^{\mathcal{S}}$ and $\beta^{\mathcal{S}}$ are the coupling constant, the anomalous dimensions and the
$\beta$-function in scheme $\mathcal{S}$, respectively ($\gamma_0$ and $\beta_0$ are scheme independent and denote 
the corresponding lowest order coefficients).
Relations (\ref{RGI1}), (\ref{RGI2}) and (\ref{RGI3}) allow us 
to compute the Z factor of the operator $\mathcal{O}$ in any scheme and at any scale we like,
once $Z^{\rm RGI}$ is known.
Therefore, the knowledge of $Z^{\rm RGI}$ is very useful for the renormalization procedure in general.
Ideally, $Z^{\rm RGI}$ depends only on the lattice spacing $a$. Computed on a finite lattice, however,
it suffers from lattice artifacts. For a precise determination it is essential to have
these discretization errors under control.

Most quantities on the lattice are computed within the so-called RI'-MOM scheme. However,
being not covariant, this scheme is not very suitable for computing the
anomalous dimensions. Therefore, we replace (\ref{RGI3}) by
\begin{equation}
Z^{\rm RGI}(a) = \Delta\,Z^{\mathcal{S}}(M=\mu_p)\, Z^{\mathcal{S}}_{\rm RI'-MOM}(M=\mu_p) \, Z^{\rm RI'-MOM}_{\rm bare}(\mu_p,a)\,.
\label{RGI4}
\end{equation}
For the intermediate scheme $\mathcal{S}$ we have chosen a momentum subtraction
scheme.
On a lattice with linear extent $L$, the scale $\mu_P$ should
fulfill the relation
\begin{equation}
1/L^2 \ll \Lambda^2_{\rm QCD} \ll \mu^2_p \ll 1/a^2\,,
\label{eq:pts3}
\end{equation}
then $Z^{\rm RGI}(a)$ would be independent of $\mu_p$ and from the resulting plateau we could read off
the corresponding final value. The formula which is used to compute the perturbative conversion factor
 $Z^\mathcal{S}_{\rm RI'-MOM}(p)$ is given  in~\cite{Gockeler:2010yr}
together with all needed coefficients of the $\beta$-function and anomalous dimensions.
We will not give them here - the reader is referred to this reference.

\section{Perturbative subtraction of order $a^2$}

As shown in~\cite{Gockeler:2010yr} the complete one-loop
subtraction of lattice artifacts results in a very weak
$p$-dependence of the $Z^{\rm RGI}$ which allows a rather
precise determination. In the absence of this procedure there is
the question whether a ``reduced'' subtraction can do a similar
job. 
It could be based on a one-loop calculation including
all possible $O(a^2)$ terms performed by the Cyprus group~\cite{Constantinou:2009tr}.
Recently, it has been demonstrated that for some selected operators and actions
the subtraction of those terms shows 
encouraging results~\cite{Alexandrou:2012mt}.

Let us denote the $O(a^2)$ part of the one-loop contribution to the renormalization constant by $Z^{(a^2)}_{\rm 1-loop}(p,a)$. Then we define the
subtracted Z factor as
\begin{eqnarray}
Z^{\rm RI'-MOM}_{\rm bare}(p,a)_{\rm MC,sub} &=& Z^{\rm RI'-MOM}_{\rm bare}(p,a)_{\rm MC} - g_\star^2 \, Z^{(a^2)}_{\rm 1-loop}(p,a)\label{eq:pts11} 
\end{eqnarray}
where $g_\star$ can be chosen to be either the bare lattice coupling $g$ or the boosted coupling $g^{}_{\rm B}$
defined by $g^{2}_{\rm B} = g^2/P(g)$, $P(g)$ is the measured plaquette at $g$.
The final renormalization group independent Z factor is then computed from (\ref{eq:pts11}) using (\ref{RGI3}),
where we expect slightly different numbers depending on the choice of coupling $g_\star$.
As suggested by the results in~\cite{Gockeler:2010yr} we choose $g_\star=g^{}_{\rm B}$.
The subtraction terms $Z^{(a^2)}_{\rm 1-loop}(p,a)$ can be calculated to a very high precision.
Therefore, the only significant errors to $Z^{\rm RI'-MOM}_{\rm bare}(p,a)_{\rm MC,sub}$ 
are due to the Monte Carlo simulations.
\begin{figure}[!htb]
  \begin{center}
     \begin{tabular}{lcr}
        \includegraphics[scale=0.56,clip=true]
         {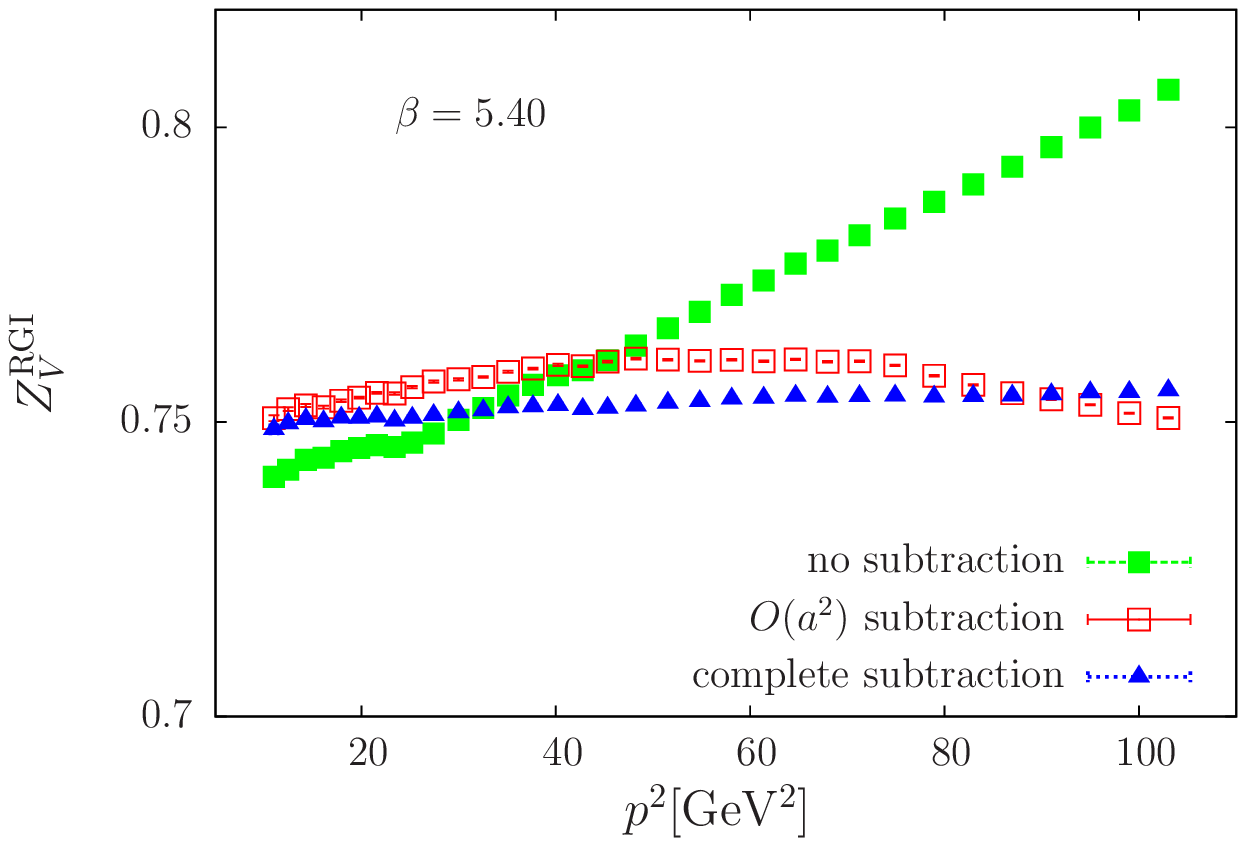}&&
       \includegraphics[scale=0.56,clip=true]
         {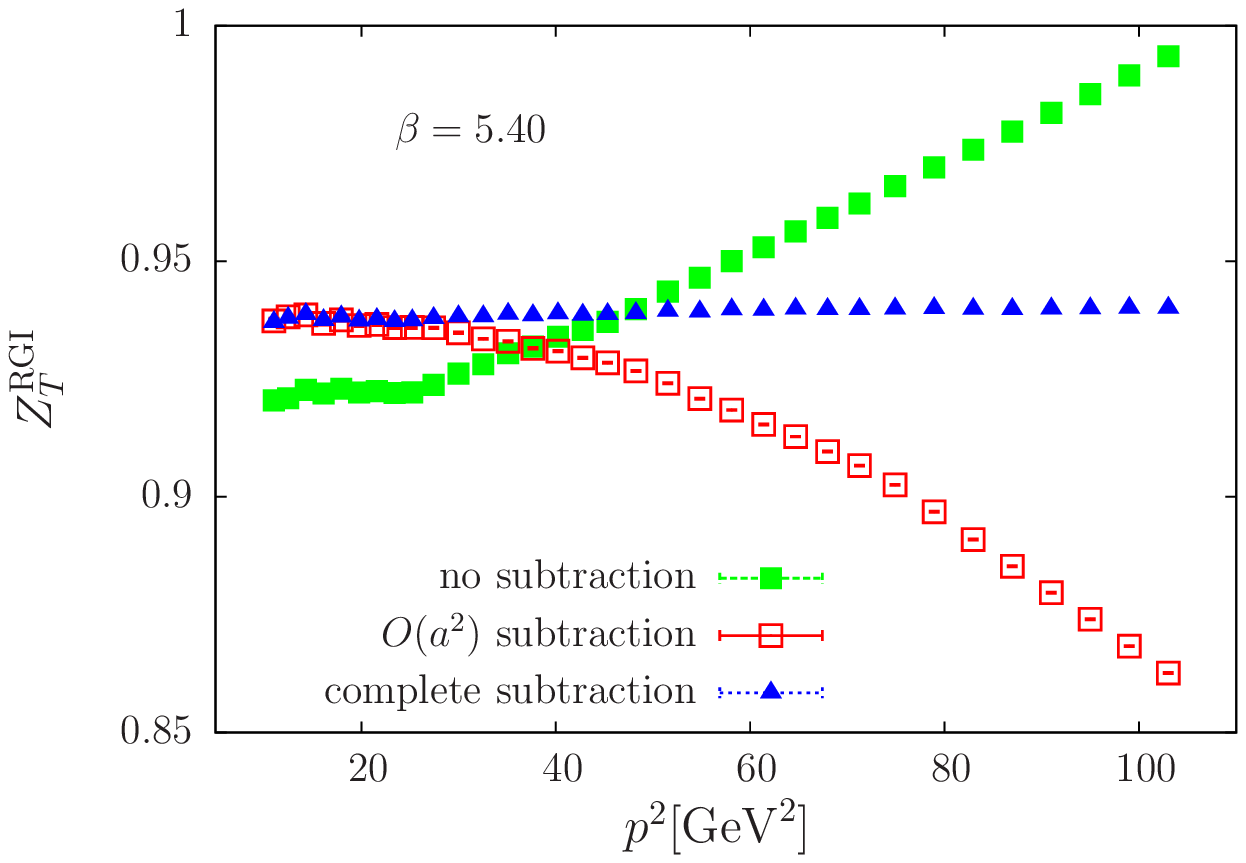}
     \end{tabular}
  \end{center}
  \caption{Unsubtracted and subtracted renormalization constants for the vector operator $\mathcal{O}^V$ (left)
and the tensor operator $\mathcal{O}^T$ (right).}
  \label{fig:ZTZva}
\end{figure}
In Figure~\ref{fig:ZTZva} we show the effect of subtraction (complete and $O(a^2)$) for the vector and tensor operators.
The complete one-loop subtraction results in a clear plateau for both
$Z^{\rm RGI}$ factors. Using the $O(a^2)$ subtraction there remains a more or less
pronounced curvature. For small $p^2$ both subtraction methods agree, as they should.

\section{Fit procedure}

Compared to the complete one-loop subtraction we expect $Z^{\rm RI'-MOM}_{\rm bare}(p,a)_{\rm MC,sub}$ as
computed from (\ref{eq:pts11}) to contain higher $p^{2n} (n \ge 2)$ terms constrained only by
hypercubic symmetry.
Therefore, we 
parametrize the subtracted data for each $\beta$ in terms of hypercubic
structures as follows (see (\ref{RGI4}))
\begin{eqnarray}
 Z^\mathcal{S}_{\rm RI'-MOM}(p) \,
			Z^{\rm RI'-MOM}_{\rm bare}(p,a)_{\rm MC,sub} &=& Z^{\rm RGI}(a)\,/\Delta Z^\mathcal{S}(p)+c_1\, a^2 \,S_2 + c_2 \,a^2 \,S_4/S_2 \nonumber \\ 
& & +\,c_3\, a^2 \,S_6/S_2^2+c_4\, a^4\, S_2^2 + c_5 \,a^4 \,S_4\label{struc1} \\
& & +\,c_6\, a^6\, S_2^3 + c_7\, a^6\, S_4 \,S_2 + c_8\, a^6 \,S_6,\nonumber
\end{eqnarray}
with $S_n=\sum_i^4 p_i^n$.
The parameters $c_1$ - $c_8$ describe the lattice artifacts. Together
with the target parameter $Z^{\rm RGI}(a)$ we have nine parameters  for this general case.
In view of the limited number of data points for each single $\beta$ value
(5.20, 5.25, 5.29, 5.40) we apply the ansatz (4.1) to all $\beta$ values
simultaneously with
\begin{equation}
Z^{\mathrm {RGI}} (a) / \Delta Z^{\mathcal S} (p) 
  \to Z_k^{\mathrm {RGI}} (a) / \Delta Z_k^{\mathcal S} (p) \,,\label{struc2}
\end{equation}
where $k$ labels the corresponding $\beta$ value.
The parameters $c_i$ are taken to be independent of $\beta$.
This enhances the ratio (number of data points)/(number of fit parameters) significantly!

Of course, there is a certain degree of freedom in the fit procedure. One choice regards 
the interval $p^2_{min} \le p^2 \le p^2_{max}$ used for the fit. Inspection of data and of the
results in~\cite{Gockeler:2010yr} suggests to use $p^2_{min} = 10\ {\rm GeV^2}$ for all $\beta$
values and all considered operators. For $p^2_{max}$ we choose the corresponding maximal available momentum. 
Another interesting point is to investigate whether the $O(a^2)$ subtraction has been 
sufficient to subtract (almost) all $p^2$ dependence. Therefore, we perform two kinds of
fits: one with all hypercubic structures under consideration $(Z_i^{\rm RGI}(a),c_1, c_2, c_3, c_4, c_5,$ $c_6, c_7, c_8)$
and one with the  $a^2$ dependence omitted $(Z_i^{\rm RGI}(a),c_4, c_5, c_6, c_7, c_8)$.

Additionally, the renormalization factors are influenced by the choice for $r_0\,\Lambda_{\rm \overline{MS}}$.
This quantity enters $\Delta\,Z^{\mathcal{S}}(M)$ in  (\ref{RGI2}) via the corresponding coupling
$g^{\mathcal{S}}(M)$ (for details see~\cite{Gockeler:2010yr}). We use two values: as standard value we take $r_0\,\Lambda_{\rm \overline{MS}}=0.73$,
suggested by~\cite{Leder:2010kz}, as a second value we choose $r_0\,\Lambda_{\rm \overline{MS}}=0.78$,
close to the result given in~\cite{Fritzsch:2012wq}.
The Sommer scale $r_0$ is chosen as $r_0=0.5 \, {\rm fm}$. The relation between the lattice spacing $a$
and the inverse lattice coupling $\beta$ is given by $r_0/a =6.050(\beta=5.20), 6.603(\beta=5.25),
6.983(\beta=5.29)$ and $8.285(\beta=5.40)$~\cite{Bali:2012}.

\section{Results}

\begin{figure}[!htb]
  \begin{center}
     \begin{tabular}{cc}
        \includegraphics[scale=0.6,clip=true]
         {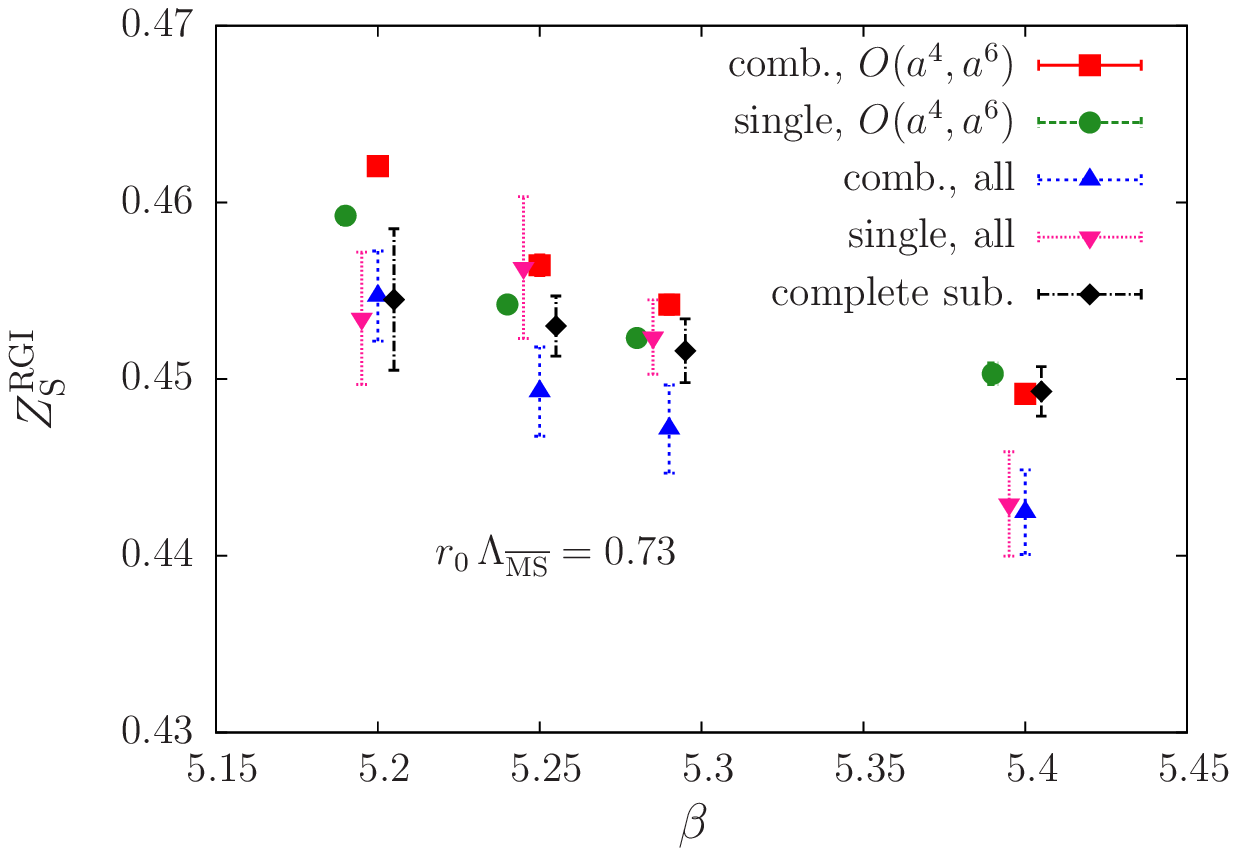}
        &
         \includegraphics[scale=0.6,clip=true]
         {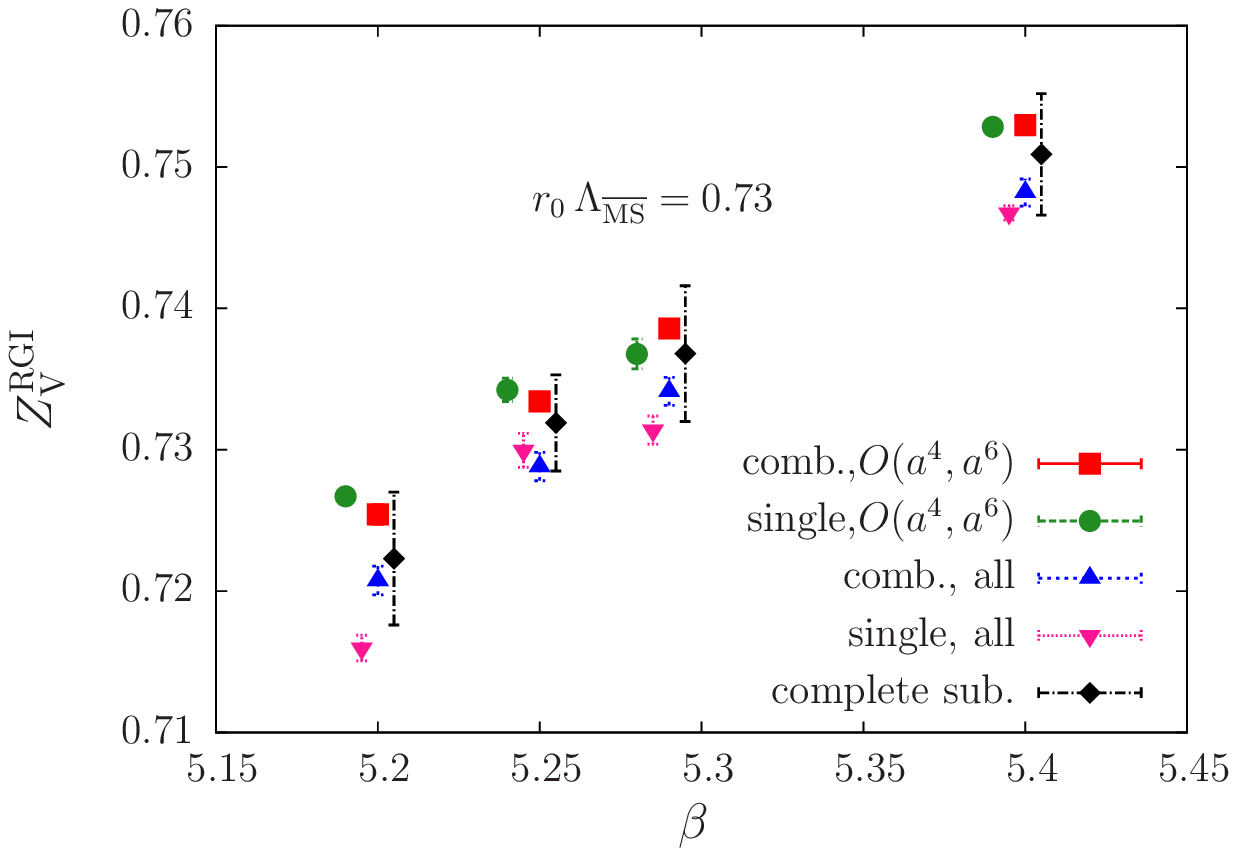}\\
         \includegraphics[scale=0.6,clip=true]
         {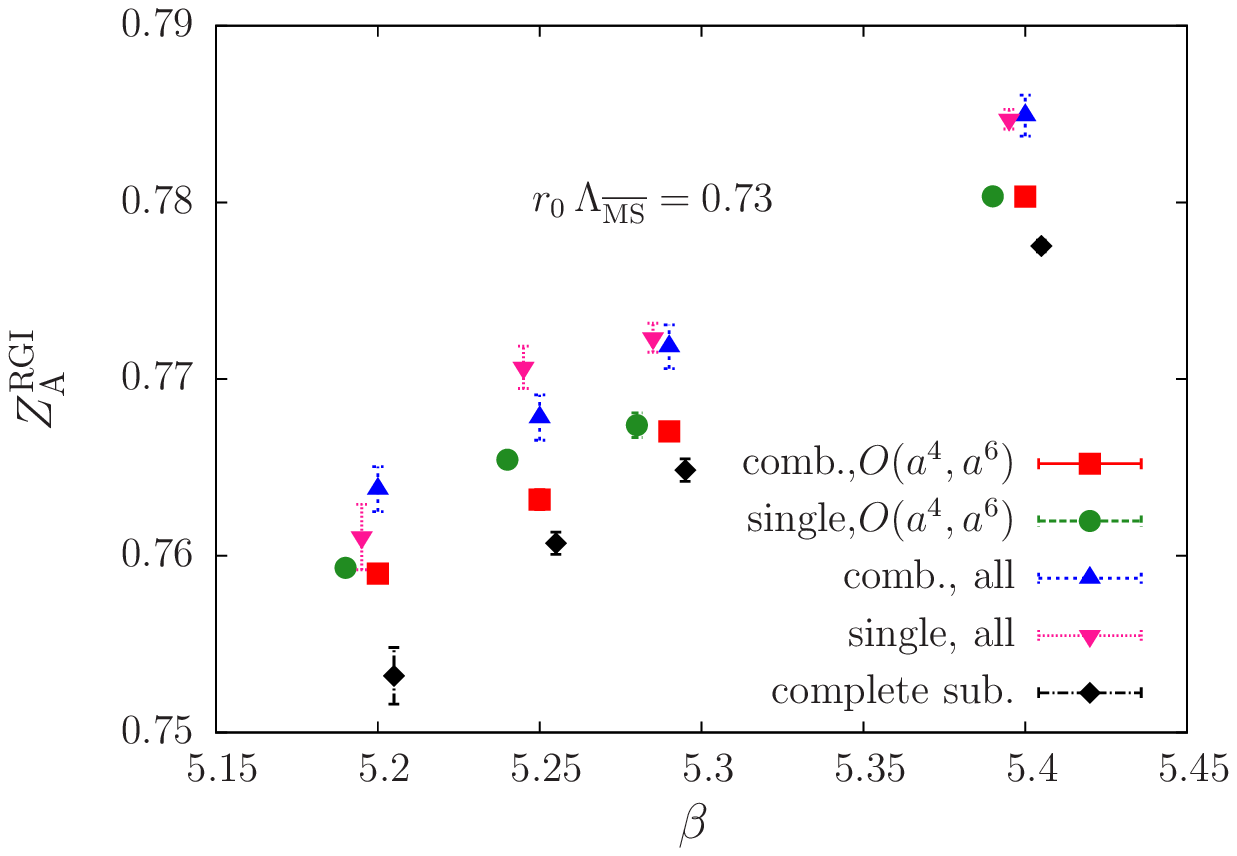}
        &
         \includegraphics[scale=0.6,clip=true]
         {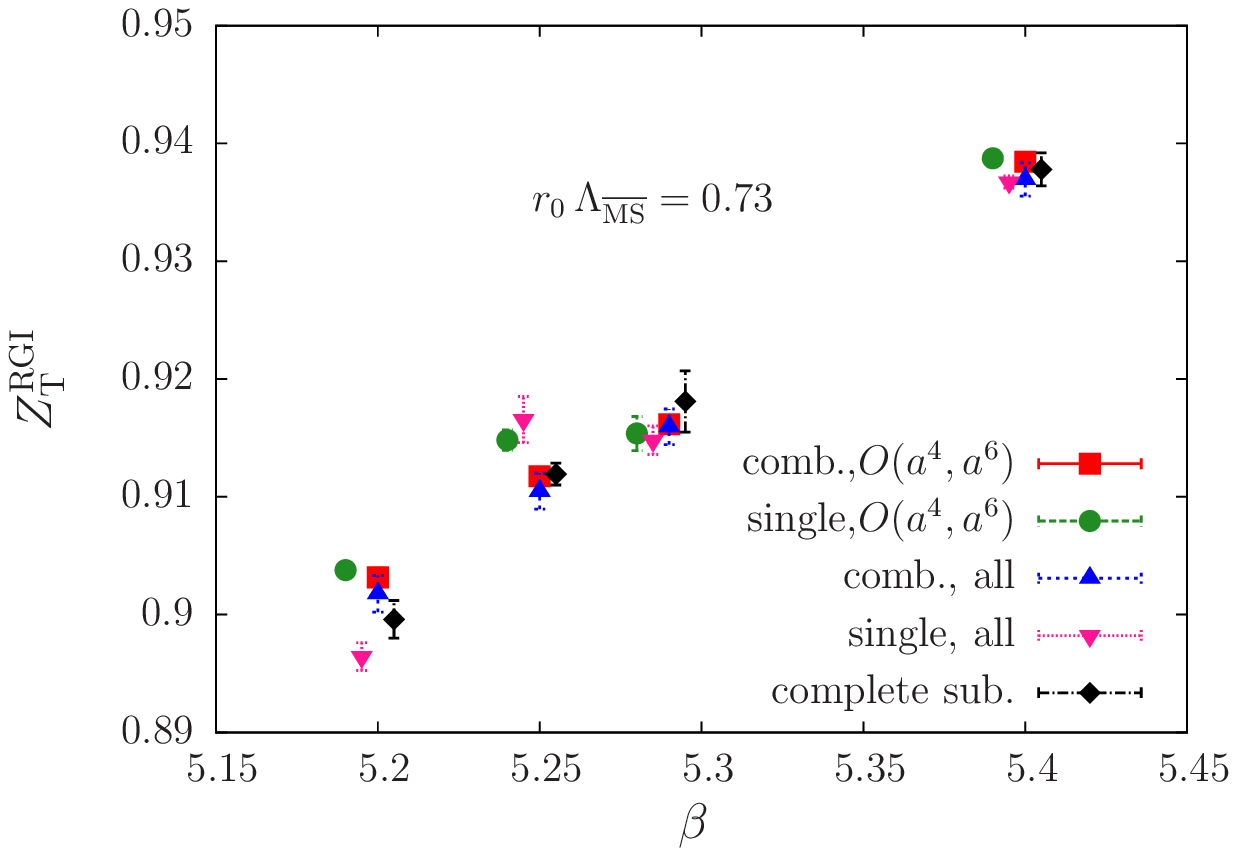}         
     \end{tabular}
  \end{center}
  \caption[]{$Z^{\rm RGI}$ for the local point operators $S, V, A$ and $T$ at $r_0\,\Lambda_{\rm \overline{MS}} = 0.73$. 
The legends denote:
``comb.'' uses the fit ansatz (\ref{struc1})+ (\ref{struc2}),
``single'' is based on (\ref{struc1}) for each $\beta$, ``all'' fits the general hypercubic structure and
``$O(a^4,a^6)$'' only possible $a^4$ and $a^6$ parts. The data points for ``complete sub.'' are obtained from the fit procedure
discussed in~\cite{Gockeler:2010yr}; they serve as reference.}
  \label{fig:ZRGI}
\end{figure}

In Figure \ref{fig:ZRGI} we show the results for our fit procedures for the choice $r_0\,\Lambda_{\rm \overline{MS}} = 0.73$.
The $Z^{\rm RGI}$ factors for the operators
$V$ and $T$ coincide within errors well
with the complete one-loop subtraction calculation. 
The renormalization factors for
$S$ and $A$ differ more significantly.
Now, we compare the different fit procedures using the perturbatively subtracted
$O(a^2)$ contributions (see (\ref{eq:pts11})).
Generally, one can state that fitting only $O(a^4,a^6)$ structures still leads to reasonable
results with smaller errors. This would mean that the subtraction of one-loop $O(a^2)$ terms
takes into account (almost) all lattice artifacts proportional to $a^2$.
In addition, 
we do not find (at least for the operators considered) a remarkable difference using either a single fit for each 
individual $\beta$ data set or a combined fit for all four $\beta$ values.

\begin{figure}[!htb]
  \begin{center}
     \begin{tabular}{lcr}
        \includegraphics[scale=0.56,clip=true]
         {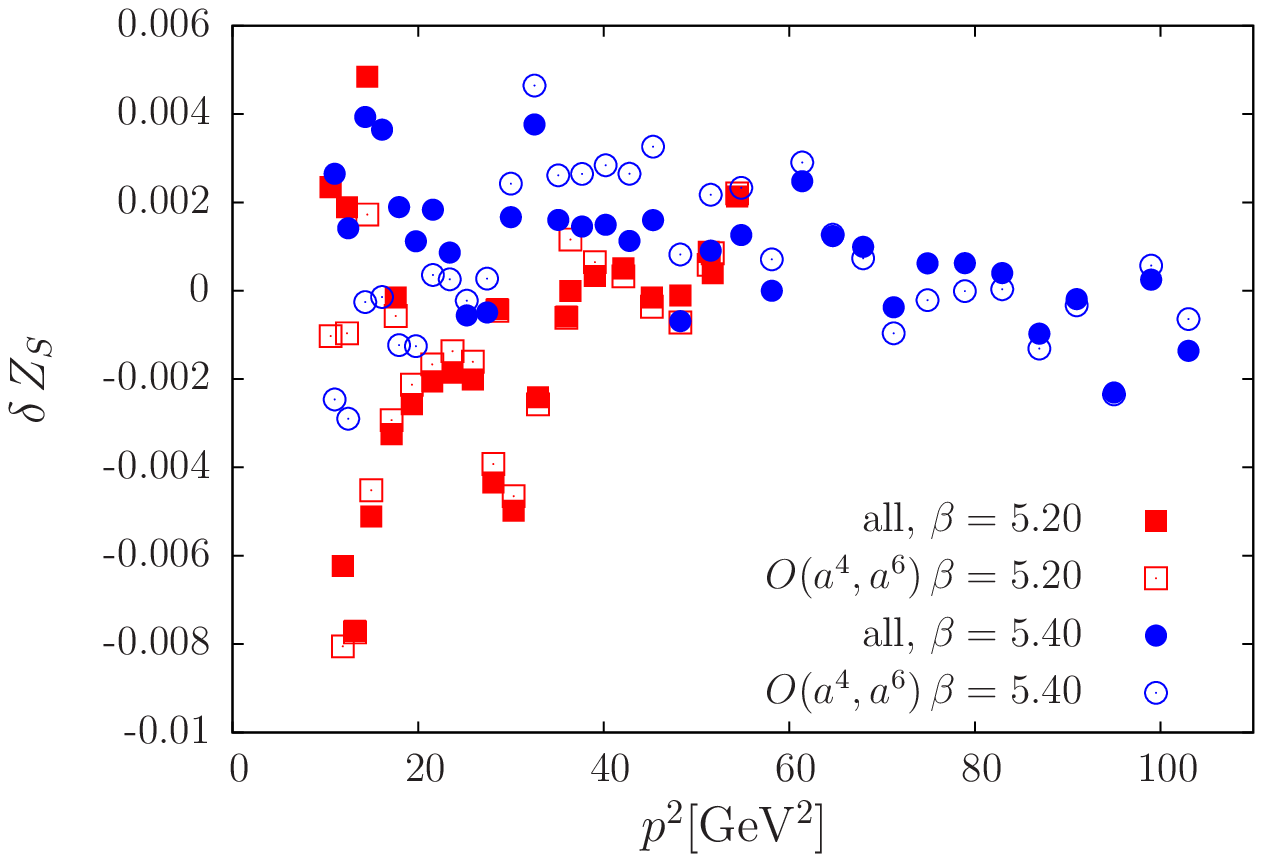}&&
       \includegraphics[scale=0.56,clip=true]
         {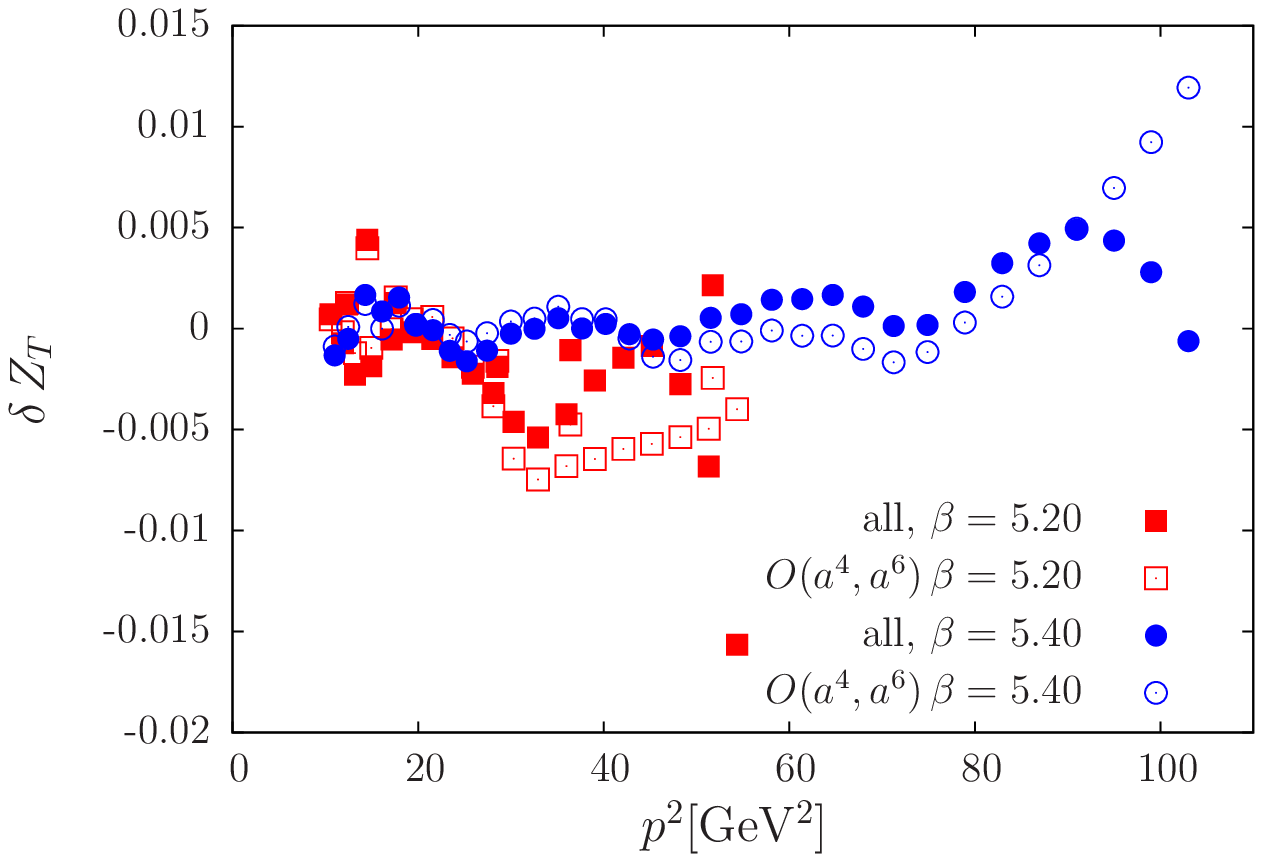}
     \end{tabular}
  \end{center}
  \caption{Relative errors $\delta Z$ for the scalar operator $\mathcal{O}^S$ (left)
and the tensor operator $\mathcal{O}^T$ (right) for $r_0\,\Lambda_{\rm \overline{MS}}=0.73$.
``all'' denotes the fit to the general hypercubic structure and
``$O(a^4,a^6)$'' only to possible $a^4$ and $a^6$ parts.}
  \label{fig:diffZ}
\end{figure}

In order to estimate the quality of the fit we compute the relative difference 
\begin{equation}
\delta Z(p) = (Z_{\rm data}(p)-Z_{\rm fit}(p))/Z_{\rm data}(p),
\end{equation}
where $Z_{\rm data}(p)$ are the data for $Z^\mathcal{S}_{\rm RI'-MOM}(p) \,Z^{\rm RI'-MOM}_{\rm bare}(p,a)_{\rm MC,sub}$. $Z_{\rm fit}(p)$ is the result of the corresponding fit. In Figure \ref{fig:diffZ} we show these
differences for the operators $\mathcal{O}^S$ and $\mathcal{O}^T$  (the other Z factors for different
$\beta$ behave similarly). One recognizes that the $\delta Z(p)$ are essentially in the per mill range. Moreover, the figures suggest that
the fit to $O(a^4,a^6)$ structures seems to be sufficient compared to fitting all structures 
in (\ref{struc1}).

\begin{table}[!htb]
\begin{center}
\begin{tabular} {|c|c| l|l|l|l|}
\hline
Op. & $r_0\,\Lambda_{\rm \overline{MS}}$ & $Z^{\rm RGI}\big|_{\beta=5.20}$ & $Z^{\rm RGI}\big|_{\beta=5.25}$ &$Z^{\rm RGI}\big|_{\beta=5.29}$ &$Z^{\rm RGI}\big|_{\beta=5.40}$ \\
 \hline
$\mathcal{O}^{\rm S}$ & $0.73$&  $  0.4592(5)(1)$  &  $  0.4542(3)(3)$  &  $  0.4523(3)(3)$  &  $  0.4503(7)(1)$   \\
$\mathcal{O}^{\rm V}$ & $0.73$&   $  0.7267(5)(3)$  &  $  0.7342(9)(5)$  &  $  0.7368(11)(7)$  &  $  0.7528(4)(5)$   \\
$\mathcal{O}^{\rm A}$ & $0.73$&  $  0.7593(4)(-11)$  &  $  0.7654(4)(-1)$  &  $  0.7674(7)(-1)$  &  $  0.7803(3)(-4)$   \\
$\mathcal{O}^{\rm T}$ & $0.73$&  $  0.9038(3)(1)$  &  $  0.9148(9)(21)$  &  $  0.9154(15)(9)$  &  $  0.9387(3)(2)$   \\
 \hline
$\mathcal{O}^{\rm S}$ & $0.78$&  $  0.4699(6)(-4)$  &  $  0.4650(6)(-2)$  &  $  0.4633(5)(-2)$  &  $  0.4579(6)(-4)$   \\
$\mathcal{O}^{\rm V}$ & $0.78$&  $  0.7265(5)(4)$  &  $  0.7340(9)(5)$  &  $  0.7366(11)(7)$  &  $  0.7527(5)(5)$   \\
$\mathcal{O}^{\rm A}$ & $0.78$&  $  0.7591(4)(-12)$  &  $  0.7652(4)(0)$  &  $  0.7672(7)(-1)$  &  $  0.7802(3)(-4)$   \\
$\mathcal{O}^{\rm T}$ & $0.78$&  $  0.8910(6)(13)$  &  $  0.9042(11)(26)$  &  $  0.9085(18)(11)$  &  $  0.9332(4)(7)$   \\
\hline
\end{tabular}
\end{center}
\caption[]{$Z^{\rm RGI}$ for the point-like operators under consideration. The results are obtained
from a fit to $O(a^4,a^6)$ structures and for each single $\beta$ data set. The fit range
in momentum space is $10\ {\rm GeV^2} \le p^2$. The results are given in the form
$value(err1)(err2)$ where $err1$ denotes the error from the nonlinear fit.
$err2$ is the shift to $value$ if the fit is performed for $8\ {\rm GeV^2} \le p^2$.
The shown numbers for $value$ correspond to the (green) full circles in Figure \ref{fig:ZRGI}.}
\label{tab:ZRGITab}
\end{table}
In Table \ref{tab:ZRGITab} we give the results 
for a fit to the higher order lattice artifact terms $O(a^4,a^6)$, where the 
$Z^{\rm RGI}$ are obtained from each single $\beta$ data set.
The results are given in the form $value(err1)(err2)$, where $err1$ is the error
of the fit parameters in the applied nonlinear
fit algorithm. $err2$ denotes the change in the  results if one uses
the fit range $8\ {\rm GeV^2} \le p^2$ which is not totally excluded.
It can serve as an indicator of systematic error.

Our fit procedure suggests that a ``reduced'' perturbative subtraction 
with a subsequent fit of $O(a^4,a^6)$ lattice artifacts leads to reliable results.
This algorithm can be used for operators with higher numbers of derivatives where a complete one-loop subtraction is not
available. However, it requires a careful investigation of the prerequisites
and the parameter choices for each new operator. Finally, one should add that
the computation of the $O(g^2a^2)$ terms in a one-loop calculation for those
operators is also challenging.
\\ \\
This investigation has been supported partly by DFG under contract SCHI 422/8-1, by the EU grant
283286 (HadronPhysics3), by SFB/TRR55 (Hadron Physics from Lattice QCD) and by RPF(Cyprus) grant NatSci/0311.


\end{document}